# Usage Policy-based CPU Sharing in Virtual Organizations

Catalin Dumitrescu
*Department of Computer Science*
*University of Chicago*
catalind@cs.uchicago.edu

Ian Foster
*Mathematics and Computer Science*
*Division, Argonne National Laboratory*
*& University of Chicago*

## Abstract

*Resource sharing within Grid collaborations usually implies specific sharing mechanisms at participating sites. Challenging policy issues can arise within virtual organizations (VOs) that integrate participants and resources spanning multiple physical institutions. Resource owners may wish to grant to one or more VOs the right to use certain resources subject to local policy and service level agreements, and each VO may then wish to use those resources subject to VO policy. Thus, we must address the question of what usage policies (UPs) should be considered for resource sharing in VOs. As a first step in addressing this question, we develop and evaluate different UP scenarios within a specialized context that mimics scientific Grids within which the resources to be shared are computers. We also present a UP architecture and define roles and functions for scheduling resources in such grid environments while satisfying resource owner policies.*

## 1. Introduction

Policy issues arise at multiple levels when sharing resources. Resource owners granting virtual organizations (VOs) the right to use certain resources want to express and enforce the usage policies (UPs) under which these resources are made available. VO representatives want to access and interpret UP statements published by resource owners. VOs typically also wish to represent and apply their own policies governing how resources aggregated from multiple resource owners are to be used. Both owners and VOs want to verify that policies are applied correctly. In this paper, we examine how UPs affect resource scheduling at both the resource owner and VO levels. We measure the impact of introducing UPs by means of two metrics: the aggregated site load (ARU) in meeting owner requirements and VOs achieved aggregated response time (ART).

Resources may include computers, storage, and networks; owners may be either individual scientists or sites; and VOs are collaborative groups, such as scientific collaborations. A VO [8] is a group of participants who seek to share resources for some common purpose. From the perspective of a single site in a Grid such as Grid3 [2], a VO corresponds to either one or several users, depending on local access policies. However, the problem is more complex than a cluster fair-share allocation problem, because each VO has different allocations under different scheduling policies at different sites and, in parallel, each VO might have different task assignment policies. This heterogeneity makes the analogy untenable when there are multiple sites and VOs.

We assume that individual resource owners negotiate service level agreements (SLAs) with each relevant VO to establish what resources are available for use by each VO. Those SLAs are based on the UP statements at each site. VOs must then aggregate resources provided by different owners to different VO purposes, and orchestrate distributed computations to use those aggregated resources efficiently. This problem encompasses challenging and interrelated UP, scheduling, and security issues. We focus in this paper on UP issues only. Specifically, we address the questions: "*What UP configuration is best suited to the Grid3 environment with many VOs and sites?*" and "*How UPs can be made available to VO schedulers and data planners for better resource utilization?*"

In addressing these questions, we build on much previous work concerning the specification and enforcement of local scheduling policies [19,23]; for negotiating service level agreements (SLAs) with remote resource sites [4,6]; and for expressing and managing VO policy [5]. We introduce the notion of UP for grid resources, measure the achieved ARU/ART under different task assignment policies, and introduce an UP infrastructure that provides support for UP and a feasible solution for the second question in Grid3 [2].

The rest of this article is as follows. We provide first a description of the scenario that we seek to address and identify the main players. In section 3 we elaborate the usage policy specifications used for scientific Grids. In section 4, we simulate and measure how well different UPs suit both user and resource owner perspectives. Section 5 contains the description of our usage policy infrastructures built in the Grid3 context [2]. Section 6 presents several related work streams, and Section 7 includes our conclusions and future plans.

## 2. Motivating Scenario

To motivate why UP-based resource sharing management is important, we consider Grid3 [2]. This system comprises numerous resources, resource owners, VOs, and resource users. Each user and resource owner participates in and may contribute resources to multiple collaborative projects that can vary widely in scale, lifetime, and formality. (The largest collaborations associated with Grid3 encompass thousands of scientists at more than one hundred institutions.) Each such project generates workloads comprising dynamic mixes of work of varying priority, some requiring the efficient aggregation of large quantities of computing and storage.

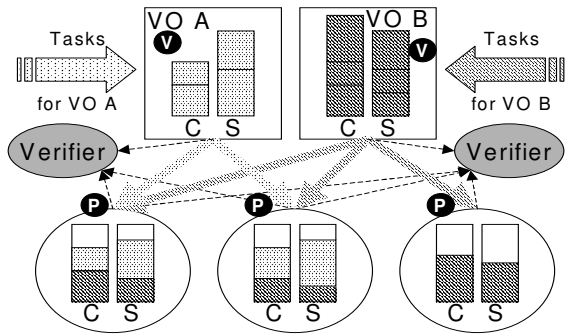

**Figure 1: Resource allocation schematic**

Figure 1 shows our model of resource allocation, in an architectural view that we describe in more detail in Section 5. In the two VOs (squares) and three sites (circles), shaded elements indicate the compute (C) and storage (S) resources allocated to each VO at each site. Sites and VOs share resources by defining *how* resource usage takes place in terms of *where*, *what*, *who*, and *when* it is allowed.

In this context, there are three dimensions in the UP space, consisting of resource provider (sites), resource consumer (VOs), and time, which are modeled as UP attributes expressed in terms of limits. Policy makers who participate in such collaboration define resource usage policies involving various levels of this space.

*UPs* represent owner statements about the policies that govern how their resources are to be allocated to resource consumers: i.e., they encode the SLAs that an owner has established with consumers. Such statements can vary in complexity from simple priority rules enforced at local resource managers to more complex descriptions that support complex SLAs and are implemented by specialized policy handling modules [17,24,25]. While diversity can be useful and may be inevitable, it is also important that resource consumers be able to interpret UPs; thus, simplicity and uniformity are to be desired.

### 2.1. Usage Policy Issues

Policy specification, enforcement, negotiation, and verification mechanisms are required at multiple levels within this environment.

*Owners* want convenient and flexible mechanisms for expressing the policies that determine how many resources are allocated to different purposes, for enforcing those policies, and for gathering information concerning resource usage. Particular concerns may include achieving the policy flexibility needed to meet different and time-varying demands, and being able to resolve disputes as to whether policies were correctly enforced.

*VOs* want to know UPs that owners impose under which resources are made available. VOs also want to interpret such SLAs to determine available resources and to guide resource allocation and task scheduling decisions. They also want to evaluate delivered service to verify that owners are respecting published UPs. VOs need to respond to resource requests from VO participants in ways that not only respect owner and VO policies but also, perhaps, maximize various performance metrics such as throughput. And VOs need to demonstrate to their participants that VO policies are followed correctly. In these latter respects, VOs also act as "owners" of the aggregate resources made available to them by sites.

### 2.2. Resource Providers and Consumers

The preceding material introduces the primary actors in the Grid3 UP environment and their interactions. Specifically, our policy framework deals with two classes of entities: resource providers and resource consumers. A physical site is a resource provider; a VO is a consumer (consuming resources provided by a site) and a provider (providing resources to users or user groups). We assume that each provider-consumer relationship is governed by an appropriate SLA, but do not address here the nature of those SLAs, an issue that can involve many trade-offs. For example, providers may prefer powerful but complex SLAs that provide flexibility in terms of how they allocate resources, while consumers may want simple SLAs that are easily interpreted.

We use an example to illustrate some issues that can arise. Assume that provider P has agreed to make R resources available to consumer C for a period of one month. How is this agreement to be interpreted? These resources might be dedicated to C or, alternatively, P might make them available to others when C is not using them. In the latter case, P might commit to preempt other users as soon as C requests them, or might commit to preempt within a certain time period. If C is allowed to acquire more than R resources when others are not using them, then this may or may not result in C's allocation being reduced later in the month. C may or may not allow reservations. The SLA between P and C should presumably capture such issues, as well (optionally) as issues such as price and acceptable use policy [17,29].

## 3. Evaluating Usage Policies

Having described the principal actors in the policy space and their requirements, we turn out attention to the question of the policies that may be used at sites. We define three specific usage policies and also a set of task assignment policies and evaluation criteria used in our experiments.

### 3.1. Usage Policies

We consider now three UPs that are likely to be deployed in real contexts such as Grid3, namely, *fixed-limit*, *extensible-limit*, and *commitment-limit*.

*Fixed-limit:* A UP statement specifies a hard upper limit on the fraction of resources $R_i$ available to a $VO_i$. A request to run a job is granted if this limit is not exceeded, and rejected otherwise. More precisely, a job requiring J resources is admitted if and only if $C_i + J \leq R_i$, where $C_i$ denotes the resources currently consumed by $VO_i$ at the site. Note that a job that is admitted will always be able to run immediately, unless the resource owner oversubscribes resources, i.e., $\Sigma_i R_i > 1$.

*Extensible-limit:* A UP statement also specifies an upper limit, but this limit is enforced only under contention. Thus, we introduce a second condition under which a job requiring J resources is admitted, namely $J \leq C_{free}$, where $C_{free}$ denotes the site's current unused resources. Note that because this policy allows VOs to consume more than their allocated resources, whether or not an admitted job can run immediately may depend on the site's preemption policy.

*Commitment-limit:* A UP statement now specifies two upper limits, an epoch limit $R_{epoch}$ and a burst limit $R_{burst}$, and for each also specifies an associated interval, $T_{epoch}$ and $T_{burst}$ respectively. A job is admitted if and only if (a) the average resource utilization for its VO is less than the corresponding $R_{epoch}$ over the preceding $T_{epoch}$, or (b) there are idle nodes and the average resource utilization for the VO is less than $R_{burst}$ over the preceding $T_{burst}$. More precisely, any jobs that verify the following algorithm are admitted.

```
# Case 1: over-used site by VO_i
if EA_i > EP_i
    reject job from VO_i
# Case 2: un-allocated site
else if Σ_k(BA_k) = 0 and BA_i + J < BP_i
    run job from VO_i
# Case 3: sub-allocated site
else if Σ_k(BA_k) + J < TOTAL and BA_i + J < BP_i
    run job from VO_i
# Case 4: over-allocated site
else if Σ_k(BA_k) = TOTAL and BA_i + J < EP_i
    schedule job from VO_i
else
    reject job from VO_i
```

with the following definitions:
- $EP_i$ = Epoch Usage Policy for $VO_i$, i.e., $R_{epoch}$
- $BP_i$ = Burst Usage Policy for $VO_i$, i.e., $R_{burst}$
- $BA_i$ = Burst Resource Usage for $VO_i$
- $EA_i$ = Epoch Resource Usage for $VO_i$
- TOTAL = upper limit allocation on the site

The following example illustrates the notation used to represent commitment-limit UPs in our work [17]:

[CPU, $Site_1$, $VO_0$, (1month, 10%), (1day, 40%)]

This policy specifies $Site_1$ will provide 10% of its computing power to $VO_0$ as long that the VO keeps a steady workload. The VO also can spike to 40% as long as free resources are available and $VO_0$ does not exceed its 10% long term allocation. If the entire allocation is consumed, $VO_0$ obtains further resources only in the next UP interval, in the next month in our example.

### 3.2. Task Assignment Strategies

The effectiveness of a specific UP may depend on the strategies used by consumers to assign tasks to nodes, and thus evaluate each UP in the face of different task assignment strategies.

De Jongh [16] distinguishes *static* scheduling policies, which do not consider any time-varying information such as load when selecting a site for execution, and *dynamic* policies, which take into account factors such as site load and site capability.

We started from two static policies, *Random* and *Round-Robin*, which assign each task to a randomly selected node and the "next" node, respectively. We extend these policies to consider only those sites at which current conditions would allow a task to be admitted. (Thus, neither extended policy is static as defined by DeJongh [16].) We also consider one dynamic policy, *Least-Used*, which assigns each task to the node that is currently the least loaded. We chose Least-Used because of its simplicity, even though it does not deliver the best performance in distributed environments, because of partial information available at any moment. Other more complex scheduling policies [16] are beyond our purpose in this paper, given our assumption that resource control in grid environments is considered distributed.

### 3.3. VO-Centric Ganglia Simulator

We have developed VO-Centric Ganglia [28], an meta-cluster simulator that can be used to study the behavior of different scheduling schemes, scheduling policies and workloads [28]. This system is derived from the Ganglia monitoring system by replacing its collectors ("gmonds") with a module that supplies simulated system information. A part of the Ganglia [18] code is then used to run the simulator code, to collect simulated states, to aggregate monitoring metrics, to maintain the history log of the system and to provide a simple interface for data accessibility.

VO-Centric Ganglia is a discrete event simulator, which means that every X seconds the simulator evaluates the state of all components in the system (jobs, queues, resource status, allocations, usages, etc). The system allows simulating a specific Grid environment, composed of a pre-specified number of sites, VOs and groups, VO planners and site managers, different task assignment policies, and usage limits. The simulator captures various costs associated with job execution, such as submission, staging, and termination. During each simulation step, various algorithms are used to adjust the state of different components in the framework.

Jobs are submitted into planner queues and then moved to site queues based on different task assignment strategies. When a job is submitted but rejected at a site (e.g., because no disk space is available or epoch shares are exhausted), it is returned to the planner queue and re-entered the planning process. If there are no available burst shares, the job is held in the planner queue and scheduled as soon as shares become available for that VO. If the site's network is not available, the job is queued for transfer and actually moved to a site when the network becomes available.

In the simulations described in this paper, VO-Centric Ganglia considers that once a job is scheduled to a site, the site runs the job even with delays. The delays can occur from site over allocation for keeping a steady workload at the site, and/or previous work that is not preempted. If no site is available for a workload, jobs are delayed from submission until a site becomes available [26].

### 3.4. Evaluation Criteria

Having defined different UPs and task assignment strategies, we need to consider how to evaluate the effectiveness of different combinations of policies and strategies in practice. We proceed as follows. We assume an oversubscribed environment in which multiple VOs can generate more work than available sites can handle. Then, we evaluate the effectiveness of different UPs by measuring both *aggregated resource utilization* (a measure of how well a UP specification meets resource owner intent) and *aggregated response time* (a measure of how well a UP specification delivers resources). Both metrics are important.

We define *aggregated resource utilization* (ARU) as the ratio of the CPU-resource actually consumed by users ($ET_i$) to the total CPU-resources available. We compute this quantity as follows:

$$ARU = \Sigma (ET_i) / (\#_{cpus} * \Delta t)$$

We define the *aggregated response time* (ART) for an entire VO [12,19] as follows, with $RT_i$ being the individual job time response:

$$ART = \Sigma_{i=1..N} RT_i / N.$$

## 4. Simulation Studies

We used VO-Centric Ganglia for simulation studies in order to evaluate alternative UP and task assignment schemes. We present results for a simulated environment comprising 10 sites (the approximate amount of available resources on Grid3 at the testing moment) with 7, 7, 7, 15, 15, 15, 15, 27, 27, and 39 CPUs, respectively, for a total of 174 CPUs. We assume six VOs, each with two workloads (described below); and implement the three UPs and three task assignment policies described earlier. We also implement a "no-limit" UP, in which site owners accept any and all tasks on a first-come/first-served basis, meaning that a particular workload might consume the entire computing capacity of a site.

The following four UP statements are a subset of those used in our experiments (the full set is available online [26]). All such statements are in effect without change throughout the simulation runtime.

(1) [CPU, Site1, VO0, (1hour, 10%), (1minute, 40%)]
(2) [CPU, Site2, VO0, (1hour, 10%), (1minute, 40%)]
(3) [CPU, Site1, VO1, (1hour, 20%), (1minute, 60%)]
(4) [CPU, Site2, VO1, (1hour, 20%), (1minute, 60%)]

These UP statements are interpreted as follows. For the fixed and extensible UPs, only the limit from the first tuple is considered, while for the commitment UP, the intervals and limits specified in both tuples are considered. We note that the time periods were chosen for ease of evaluation in VO-Centric Ganglia (discrete simulator using real time yet) and are significantly smaller than the real values we would expect to use in real deployments, where the overall duration of a policy might be on the order of weeks or months, and the burst period might be on the order of hours to days.

### 4.1. Workloads

We use synthetic workloads to evaluate our usage policies. Each workload is composed of jobs, each corresponding to a certain amount of work and with precedence constraints determining the order in which jobs can be executed. Jobs arrive, are executed, and leave the system according to a Poisson distribution. Because in our simulations we consider an environment with several VOs, an important factor is synchronization among workloads. We consider two cases: all workloads start at the same moment in time (synchronized), and workloads start at different moments in times (un-synchronized) [16,17].

We defined 6x2x4 simple workloads, each of which overlays work entering the grid for six VOs. Workloads are associated with VO groups, and each group had associated four burst workloads scheduled at predefined time intervals (both synchronous and un-synchronous cases). In addition, the simulator takes as input a set of files associated with each job and inter-file dependencies, which captures the inter-job dependencies.

Each workload consisted of 440 jobs, as shown in Table 1. These synthetic workloads are designed to imitate various workloads running on Grid3 in terms of CPU usage [2].

**Table 1: Grid-wide workload summary**

| VO | Workload | #jobs | Avg. Job Duration |
|---|---|---|---|
| 0 | 0 | 80 | 200 sec |
| 0 | 1 | 100 | 300 sec |
| 1 | 0 | 120 | 150 sec |
| 1 | 1 | 140 | 250 sec |

In all cases, the simulation period was one hour and the measurement interval was 30 seconds. Job durations and inter-arrival times were generated by Poisson and Gaussian distributions, respectively. Two of the six composite workloads are depicted in Figure 2. The job completion times shown above are idealized; the start points of the workloads from 0 will match the simulated charts but the finish times on the simulator charts may be delayed from this "ideal."

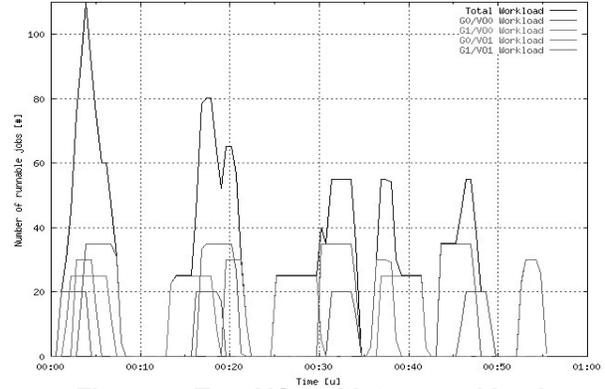
**Figure 2: Two VOs with two workloads**

### 4.2. Results

We present the results achieved for the three task assignment policies and the four UPs. (Additional results are in a separate report [26]). Tables 2 and 3 summarize results for the synchronized workloads, and Tables 4 and 5 for the un-synchronized workloads. As a side note, seconds can be interpreted as any time unit and only the length of all experiments and Grid3 resource usage constrained as from using a different time length.

**Table 2: ARU, synchronized**

| Policy/UP | No-limit | Fix-limit | Ext-limit | Cm-limit |
|---|---|---|---|---|
| *Random* | 0.72 | 0.75 | 0.69 | 0.78 |
| *Round Robin* | 0.70 | 0.65 | 0.75 | 0.77 |
| *Least Used* | 0.69 | 0.80 | 0.81 | 0.79 |

**Table 3: ART, synchronized**

| Policy/UP | No limit | Fix-limit | Ext-limit | Cm-limit |
|---|---|---|---|---|
| *Random* | 10.64 | 19.25 | 12.83 | 15.83 |
| *Round Robin* | 11.09 | 19.39 | 11.32 | 15.52 |
| *Least Used* | 13.25 | 15.14 | 15.06 | 16.02 |

We note first that Tables 3 and 5 indicate that the best user performance is achieved in the absence of any UP constraints (column *No limit*). This behavior is what we should expect: UPs function by rejecting jobs that could otherwise run. We see also that in this situation, *Random* performs marginally better than *Round Robin* and much better than *Least Used* for the synchronized workloads, while in the unsynchronized case, *Round Robin* is clearly superior. Further work is required to account for these differences.

Considering next the behavior of different task assignment policies in the presence of UPs other than *No Limit*, we see that from the perspective of response time, the extensible limit task assignment provides the best aggregated response time. An explanation for this result is that whenever resources are available for a VO at a site, they are grabbed without any restriction.

Table 3 shows that, for synchronized workloads, *Random Assignment* without any UP limit provides the best response time. Table 2 shows that *Commitment* provides almost the best approach for the resource owners to enforce different local preferences and priorities.

**Table 4: ARU, un-synchronized**

| Policy/UP | No-limit | Fix-limit | Ext-limit | Cm-limit |
|---|---|---|---|---|
| Random | 0.70 | 0.67 | 0.70 | 0.69 |
| Round Robin | 0.69 | 0.65 | 0.71 | 0.65 |
| Least Used | 0.69 | 0.64 | 0.72 | 0.64 |

**Table 5: ART, un-synchronized**

| Policy/UP | No limit | Fix-limit | Ext-limit | Cm-limit |
|---|---|---|---|---|
| Random | 10.3 | 12.59 | 10.64 | 13.35 |
| Round Robin | 7.78 | 14.82 | 9.34 | 12.35 |
| Least Used | 10.57 | 13.68 | 11.37 | 12.59 |

Looking next at the unsynchronized workload execution results, we see the Round Robin policy without UP limits provides the best response time, while Commitment again provides the highest site utilization. However, while the un-synchronized workloads result in better user performance, aggregated site usage is lower by almost 10% for Fixed Limit and Extensible Limit UPs, while 10% higher for Commitment UP.

## 5. Experimental Studies

We have constructed a partial prototype framework to investigate the feasibility of integrating policies within Grid3 [2]. The components of the envisaged framework are described in Figure 3 [30].

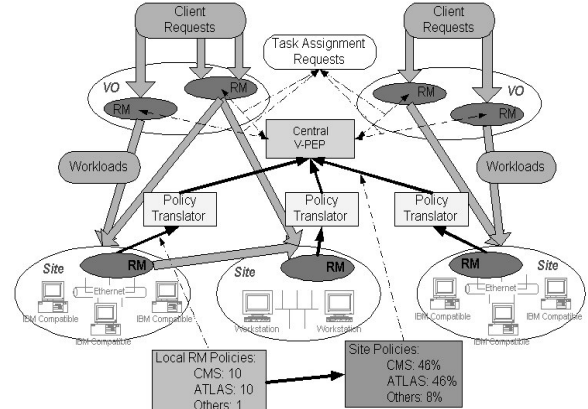

**Figure 3: Revised architecture (one V-PEP)**

*Resource Managers* (RMs) represent the layer at which most resource managers (e.g., cluster schedulers [19,20], network reservation mechanisms [22]) operate. RMs are assumed to have basic mechanisms for job prioritization and resource acquisition.

*Policy Translators* (PTs) usually take as input some RM low level priority description and publish further after a specific translation operation.

*VO policy enforcement point* (V-PEP) make decisions on a per-job basis to enforce UP regarding both VO specifications and site specifications. V-PEP is invoked whenever a VO planner makes job planning and scheduling decisions.

To verify the performance of this approach in meeting the simulation results presented in Section 4, we used two submission points that incorporate knowledge about each workload, site UPs, and monitoring elements for task assignment and job selection algorithms [30]. Results gathered from these experiments are presented in Tables 6 and 7, similarly to the ones in Section 5. All workload runs were done over a subset of Grid3 resources and took in account the actual local preferences [2]. Resources already busy were not counted in computing the following statistics.

**Table 6: ARU, synchronized**

| Policy/UP | No-limit | Fix-limit | Ext-limit | Cm-limit |
|---|---|---|---|---|
| Random | 0.40 | 0.32 | 0.36 | 0.54 |
| Round Robin | 0.42 | 0.46 | 0.44 | 0.65 |
| Least Used | 0.28 | 0.20 | 0.18 | 0.30 |

**Table 7: ART, synchronized**

| Policy/UP | No-limit | Fix-limit | Ext-limit | Cm-limit |
|---|---|---|---|---|
| Random | 69.98 | 83.62 | 106.48 | 82.66 |
| Round Robin | 66.78 | 73.80 | 91.06 | 78.43 |
| Least Used | 75.71 | 81.89 | 102.44 | 89.56 |

The results for the un-synchronized workloads over Grid3 are presented in Tables 8 and 9.

**Table 8: ARU, un-synchronized**

| Policy/UP | No-limit | Fix-limit | Ext-limit | Cm-limit |
|---|---|---|---|---|
| Random | 0.27 | 0.30 | 0.40 | 0.41 |
| Round Robin | 0.31 | 0.29 | 0.43 | 0.55 |
| Least Used | 0.23 | 0.16 | 0.25 | 0.33 |

**Table 9: ART, un-synchronized**

| Policy/UP | No limit | Fix-limit | Ext-limit | Cm-limit |
|---|---|---|---|---|
| Random | 66.34 | 81.02 | 90.83 | 81.83 |
| Round Robin | 62.05 | 80.83 | 73.78 | 79.71 |
| Least Used | 71.98 | 75.91 | 95.65 | 84.00 |

First, the *Round Robin* task assignment policy performs the best in almost all cases with respect to job response time. Second, for the aggregated site usages, *Commitment* does perform as well as in the simulated case. We note that latencies incurred in submitting jobs to site schedulers, and the subsequent scheduling delays, are an important factor, which is not captured by our VO-Centric Ganglia simulator [28].

In addition, we have considered only a centralized enforcement point, in other words a single point of job assignment decision. In future work, we will focus on multiple enforcement points that do not have access to global usage and UP information. Lastly, at the time of the experiments, the local workloads were constant and insignificant with the amount of work submitted through our framework. In addition, we excluded all the busy CPUs from our computations for ARU and ART.

## 6. Related Work

Fair share scheduling strategies seek to control the distribution of resources to processes so as to allow greater predictability in process execution. These strategies were first introduced in the early 1980s in the context of mainframe batch and timeshared systems and were subsequently applied to Unix systems [12,13]. We exploit these techniques in our work.

Author et al. investigate the question of scheduling tasks according to a user-centric value metric – called utility [27]. Sites sell the service of executing tasks instead of raw resources. The entire framework is centered on selling and buying services, with users paying more for better services and sites paying penalties when they fail to honor the agreed commitments. The site policies are focused on finding winning bids and schedule resource accordingly. This approach is different from our work here, as a more abstract form of resources is committed under different SLAs.

The Maui scheduler [23] is an external job scheduler for use on clusters and supercomputers. It operates as a policy engine for controlling resource allocations to jobs while concurrently optimizing the use of managed resources. Maui operates by guiding the scheduling decisions of other cluster managers. It manipulates jobs, nodes, reservations, QoS structures, policies, and composite objects. Maui is the closest work to what we described in this paper, while it is mainly a cluster scheduler providing a flexible mechanism for scheduling policy specification and enforcement at the cluster level.

LSF is a resource management product that schedules, monitors, and analyses the workload for a set of computers. LSF supports sequential and parallel applications running as interactive and batch jobs. LSF is a loosely coupled cluster solution for heterogeneous systems, with several scheduling strategies available, including Job Priority Based, Deadline Constraints, Exclusive, Preemptive, and Fair-share. Multiple LSF scheduling policies can be invoked in parallel for different sub-clusters [21]. Both Maui and LSF represent cluster level schedulers and deal with concepts such us queues, users, groups, nodes, and scheduling strategies. However, our problem is more complex due to the heterogeneity of the considered environment – many sites and VOs.

## 7. Conclusions

We have presented and evaluated an approach to representing and managing resource allocation policies in a multi-site, multi-VO environment. We present results in three distinct areas. Firstly, we experimented with UP scenarios. Secondly, we were interested to present a few simple and common UP examples that arise in grid environments, and to gather experience for other UP scenarios. Thirdly, we presented architecture for UP-based scheduling in the Grid3 environment. The main gains for Grid3 are additional usage information that gives grid planners, such as Euryale, Pegasus, or Sphinx [29], hints about what sites to consider for job placement; and time-based entitlement to resources, VOs being guaranteed under different FS policies that they get resources when they need them instead of maintaining constant workloads.

There are still problems and technical details not fully explored in this paper. For example, our analysis did not consider the case of cluster administrators that *over-subscribe* local resources, in the sense of a local policy that states that 40% of the local CPU power is available to VO1 and 80% is available to VO2. A second issue not discussed in this paper is the hierarchic grouping and allocation of resources. Generally, VOs will group their users under different schemes. We are interested to use the same policy specification for such specifications, and come with some hierarchies of usage policies to encode this. While this is an important problem in Grid3 context, we leave it as an open problem at the current stage.

**Acknowledgements:** We thank Robert Gardner, Ruth Pordes, Harvey Newman, Michael Wilde, Matei Ripeanu, and Rick Cavanaugh for insights, discussions, and support, and Asit Dan for shaping the motivating scenario. We also thank the Grid3 project. This work was supported in part by the NSF Information Technology Research GriPhyN project, under contract ITR-0086044.

## Bibliography

[1] Avery, P. and Foster, I. The GriPhyN Project: Towards Petascale Virtual Data Grids, 2001, www.griphyn.org.

[2] Grid2003 Project, "The Grid2003 Production Grid: Principles and Practice", *Proc 13<sup>th</sup> IEEE Intl. Symposium on High Performance Distributed Computing, 2004*.

[3] Chervenak, A., Foster, I., Kesselman, C., Salisbury, C. and Tuecke, S. The Data Grid: Towards an Architecture for the Distributed Management and Analysis of Large Scientific Data Sets. *J. Network and Computer Applications* (23), 187-200, 2001.

[4] Czajkowski, K., Foster, I., Kesselman, C., Sander, V. and Tuecke, S., SNAP: A Protocol for Negotiating Service Level Agreements and Coordinating Resource Management in Distributed Systems. in *8<sup>th</sup> Workshop on Job Scheduling Strategies for Parallel Processing*, '02.

[5] Pearlman, L., Welch, V., Foster, I., Kesselman, C. and Tuecke, S., A Community Authorization Service for Group Collaboration. in *IEEE 3<sup>rd</sup> International Workshop on Policies for Distributed Systems and Networks*, 2002.

[6] *WSLA Language Specification*, Version 1.0, IBM Corporation, 2003.

[7] Ranganathan, K. and Foster, I., Decoupling Computation and Data Scheduling in Distributed Data Intensive Applications. in *International Symposium for High Performance Distributed Computing*, Edinburgh, UK, 2002.

[8] Foster, I., Kesselman, C., and Tuecke, S., The Anatomy of the Grid. in *International Supercomputing Applications*, 2001.

[9] Schopf, J., and Nitzberg, B., Grids: The Top Ten Questions. in *Special Issue of Scientific Programming on Grid Computing*, 2002.

[10] Zhao, T., and Karamcheti, V., *Expressing and Enforcing Distributed Resource Agreements*. Department of Computer Science, Courant Institute of Mathematical Sciences, New York University, 2000.

[11] Lupu, E., *A Role-based Framework for Distributed Systems Management*, PhD Dissertation, Imperial College of Science, Technology and Medicine, University of London, Department of Computing, 1998.

[12] Henry, G.J., "*A Fair Share Scheduler*", AT&T Bell Laboratory Technical Journal, Vol. 63, No. 8, Oct 1984.

[13] Kay, J., and Lauder, P., *A Fair Share Scheduler*, University of Sydney and AT&T Bell Labs, 1988.

[14] Verma, D.C., *Policy Based Networking, Architecture and Algorithm*, New Riders Publishing, Nov. 2000.

[15] Kosiur, D. *Understanding Policy Based Networking*", Wiley Computer Publishing, 2001.

[16] De Jongh, J. F. C. M., *Share Scheduling in Distributed Systems*, Delft Technical University, 2002.

[17] Dumitrescu, C., Wilde, M., and Foster, I., "Policy-based CPU Scheduling in VOs", GriPhyN/iVDGL Technical Report, 2003-29.

[18] Massie, M., Chun, B., and Culler, D., The Ganglia Distributed Monitoring: Design, Implementation, and Experience. in Parallel Computing, May 2004.

[19] Condor Project, A Resource Manager for High Throughput Computing, Software Project, The University of Wisconsin, www.cs.wisc.edu/condor.

[20] OpenPBS Project, A Batching Queuing System, Software Project, Altair Grid Technologies, LLC, www.openpbs.org.

[21] LSF Administrator's Guide, Version 4.1, Platform Computing Corporation, February 2001.

[22] Foster, I., Fidler, M., Roy, A., Sander, V., and Winkler, L., *End-to-End Quality of Service for High-end Applications*. in Computer Communications, 27(14):1375-1388, 2004.

[23] MAUI, Maui Scheduler, Center for HPC Cluster Resource Management and Scheduling, www.supercluster.org/maui.

[24] RFC3060, *Policy Core Information Model -- Version 1 Specification*, www.faqs.org/rfcs/rfc3060.html.

[25] RFC3198, *Terminology for Policy - Based Management*, www.faqs.org/rfcs/rfc3198.html.

[26] Dumitrescu, C., Policy-based Resource Allocation Tools", people.cs.uchicago.edu/~cldumitr/Experiments.

[27] Irwin D., Chase J., and Grit L., Balancing Risk and Reward in Market-Based Task Scheduling. in *the Thirteenth International Symposium on High Performance Distributed Computing*, HPDC-13, 2004.

[28] Foster, I. and Dumitrescu, C., "VO-Centric Ganglia Simulator", GriPhyN/iVDGL Technical Report, '04-31.

[29] In, J., and Avery, P., Policy Based Scheduling for Simple Quality of Service in Grid Computing. in *International Parallel & Distributed Processing Symposium (IPDPS)*, Santa Fe, New Mexico, April '04.

[30] Dumitrescu C., Wilde M., and Foster I., "Usage Policy at the Site Level in Grid3", GriPhyN/iVDGL Technical Report, '04-71.